\pdfoutput=1
\documentclass[sigconf]{acmart}

\usepackage{multirow}
\usepackage{subcaption}

\setcopyright{none}
\renewcommand\footnotetextcopyrightpermission[1]{} 
\begin{document}

\title{Towards Serverless Optimization with In-place Scaling}

\author{Vincent Hsieh}
\affiliation{%
  \institution{National Taiwan University}
  \city{Taipei}
  \country{Taiwan}
}
\email{0426vincent@ntu.im}

\author{Jerry Chou}
\affiliation{
  \institution{National Tsing Hua University}
  \city{Hsinchu}
  \country{Taiwan}
}
\email{jchou@lsalab.cs.nthu.edu.tw}

\begin{abstract}
Serverless computing has gained popularity due to its cost efficiency, ease of deployment, and enhanced scalability. However, in serverless environments, servers are initiated only after receiving a request, leading to increased response times. This delay is commonly known as the ``cold start'' problem. In this study, we explore the in-place scaling feature released in Kubernetes v1.27 and examine its impact on serverless computing. Our experimental results reveal improvements in request latency, with reductions ranging from 1.16 to 18.15 times across various workloads when compared to traditional cold policy.
\end{abstract}

\keywords{Serverless Computing, Cloud Computing, Scaling, Kubernetes}

\maketitle
\pagestyle{empty}
\section{Introduction}
Serverless computing represents a novel paradigm in cloud computing that has been widely supported across both commercial cloud providers \cite{aws-lambda,azure-functions,gcf} and open-source platforms \cite{faas,knative,whisk,0lambda,fission}. It allows developers to concentrate on the core logic of their applications while abstracting away the complexities of deployment management. This model has been effectively utilized in diverse applications ranging from video processing and analytics \cite{sprocket,videos,llama}, machine learning/deep learning \cite{distributed-ml,batch,dl}, linear algebra \cite{numpywren}, and high-performance computing \cite{Mashup}.

Despite the advantages of serverless computing, there are some trade-offs. Unlike serverful applications, where servers are continuously running, servers begin the startup process, including resource allocation, code downloading, and runtime environment setup, only after receiving a request. This step introduces additional latency, commonly referred to as ``cold start''. Moreover, the unpredictability of request frequency makes anticipating and pre-launching function instances to mitigate delays more challenging. Substantial research has been conducted to reduce this latency \cite{faasnap,catalyzer,seuss,sock,faascache}. Observations from Microsoft Azure Functions \cite{serverlesswild} indicate that over 50\% of functions execute in less than one second, indicating that even brief latency can still critically impact response times. One common strategy to mitigate this issue is to keep functions in a warm state for a certain period (usually 5 to 10 minutes, varying with the workload or cloud service providers). This approach enables subsequent requests within the time frame to bypass the cold start phase and be processed immediately. However, the downside of this method is that it reserves computing resources for the warm functions, which could otherwise be allocated to other active functions in need.

In this work, we analyzed the alpha feature ``In-place Resource Resize for Kubernetes Pods'' (referred to as ``in-place scaling'' in the following context) released in Kubernetes 1.27 \cite{inplace}. Prior to this feature, vertical scaling in Kubernetes required restarting the instance since container resources are immutable after launch. However, restarting a pod in serverless computing is undesirable as it triggers additional cold start processes, significantly increasing the handling time of a request. The in-place scaling feature is a game changer, as it allows the instance to scale up and down resources without the need for a restart, which is beneficial in both serverful and serverless computing. Our study primarily focused on scaling the CPU, as it directly affects response time. Reducing memory may trigger Out Of Memory (OOM) issues, which we plan to investigate in the future. We assessed the overhead of the in-place scaling both with and without workloads present. Additionally, we integrated this feature into Knative and compared the outcomes with existing cold and warm start policies.

The remainder of this paper is organized as follows: Section \ref{sec:bg} provides the background and motivation behind our research, laying the groundwork for the subsequent analysis. Section \ref{sec:compare} offers a comparative study of three distinct scaling policies in serverless computing. In Section \ref{sec:exp}, we detail the experimental setup and methodology used to evaluate the in-place scaling feature. Section \ref{sec:related} reviews the related work, and Section \ref{sec:conclusion} concludes the paper, discussing potential directions for future research.

\section{Background and Motivation}
\label{sec:bg}
Kubernetes \cite{k8s}, recognized as a state-of-the-art, open-source container orchestration system, offers capabilities such as autoscaling, load balancing, and service discovery. It supports two primary scaling strategies: horizontal and vertical. Horizontal scaling can launch new function instances (scaling out) or remove unneeded ones (scaling in), based on metrics such as CPU, memory, or other custom metrics, to handle varying workloads \cite{hpa}.

In contrast, vertical scaling, or scaling up and down, adjusts the resources allocated to existing workloads in response to changing demands. Vertical scaling benefits several factors, including input size, parameters, workload complexity, network traffic, and latency sensitivity. For instance, tasks with larger inputs or high latency sensitivity should be allocated more resources, and the opposite for less demanding tasks. Moreover, chain functions \cite{sand} present another scenario where vertical scaling is beneficial. Consider a data processing pipeline with a sequence of functions: Data Ingestion, Data Cleaning, Data Transformation, Data Analysis, and Data Output. Each function in this chain performs a specific task within the workflow and passes its output to the next function. Vertical scaling can be applied to allocate additional resources to functions handling more complex tasks, ensuring that the entire chain operates efficiently and meets the desired performance metrics.

However, there are two challenges in achieving this setting in Kubernetes. Firstly, Kubernetes allocates pods for deployment with uniform computing resources, meaning that instances under the same deployment receive identical resource allocations, irrespective of varying external factors such as input size. Secondly, while Kubernetes employs the Vertical Pod Autoscaler (VPA) to predict resource needs based on both current and historical resource usage \cite{vpa}, before the introduction of the in-place scaling feature, applying the newly estimated resources required a restart. This process, involving service termination and recreation, can be disruptive. It has been suggested that Kubernetes' VPA will transition to an in-place scaling mechanism once it becomes available \cite{vpa}.

The allocation of CPU and memory resources to a container is a critical factor influencing performance. In AWS Lambda \cite{lambda-config}, the amount of CPU power provided is proportional to the configured memory, meaning that increasing the memory allocation directly boosts the available CPU power. Google Cloud Function allows users the flexibility to set vCPU resources. If the vCPU is not explicitly configured, it is allocated based on the memory settings \cite{0gcfc}. In Kubernetes, CPU resources are measured in units, where 1 CPU unit, equivalent to 1000 milliCPU (1000m), corresponds to either one physical CPU core or one virtual core, depending on the nature of the node. Kubernetes uses CPU requests as a weighting mechanism in scenarios of heavy system resource contention to determine the allocation of CPU time among workloads. These CPU requests are then translated into CPU shares managed by the Completely Fair Scheduler (CFS). For example, if two applications have CPU requests specified as 100m and 50m, respectively, in a scenario where CPU resources are fully utilized on the node, the application with the 100m request would be entitled to two-thirds of the CPU time, while the application with the 50m request would receive the remaining one-third. 

Most services have relied on horizontal scaling to manage workloads due to challenges in accurately predicting the appropriate amount of resources for vertical scaling. Many resource managers \cite{yarn,mesos,omega} require the specification of resources before launching applications. However, Delimitrou et al. show that 70\% of workloads oversubscribe the required resources to handle requests \cite{quasar}, indicating the difficulty of specifying and allocating the correct resources. With the release of the in-place scaling feature, Kubernetes now has the potential to effectively combine vertical and horizontal scaling. This offers a more comprehensive approach to workload management and enables more flexible and efficient resource utilization.

\begin{figure}
  \includegraphics[width=\columnwidth,keepaspectratio]{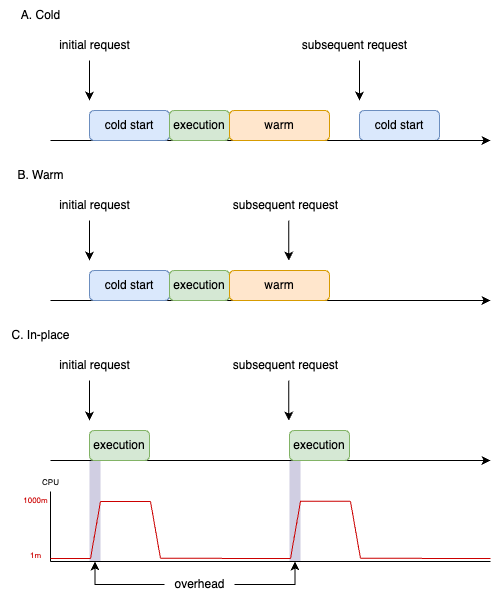}
  \caption{Different Phases of Serverless Task Execution}
  \label{fig:1}
\end{figure}

\section{Cold, Warm, In-place}
\label{sec:compare}
Figure \ref{fig:1} illustrates the behaviors of three distinct scheduling policies in serverless computing: A) \textbf{Cold}, B) \textbf{Warm}, and C) \textbf{In-place}. The primary distinction between the cold and warm policies lies in the arrival time of subsequent requests. Under the cold policy, the next request arrives either while the server is busy processing other requests or after it has been shut down. In both scenarios, no available handler is present, prompting the scheduler to launch a new function instance. This process requires a full restart and incurs the cold start latency. In contrast, the warm policy is applicable when a request arrives while the handler is still active but idle. In this warm state, the scheduler can immediately assign the request to the available handler, allowing for processing in a manner similar to a continuously running server. In our work, the in-place policy represents a state where the instance remains active on the node with minimal CPU allocation. When a request arrives, the scheduler scales up the CPU to meet demands and scales it down after task completion. The only overhead occurs during scaling up; the scheduler will redirect the request immediately after dispatching the updated configuration. As a result, the function instance serves with a small CPU allocation for a short period, adding a slight overhead. 

The introduction of the in-place scaling brings three advantages to serverless computing:

\begin{enumerate}
  \item \textbf{Reduced Latency:} By eliminating the restarts typically associated with vertical scaling, in-place scaling significantly reduces the latency that would otherwise be introduced during restart. This improved responsiveness is especially beneficial in serverless architectures, where even minimal delays can impact performance and user experience.
  \item \textbf{Enhanced Resource Availability:} Maintaining an active, low-utilization state on the node ensures that resources are conserved and can be dynamically allocated based on incoming requests. This approach prevents resource wastage and reduces the likelihood of contention, allowing for more efficient allocation of resources to active functions as needed.
  \item \textbf{Fine-grained Resource Control:} In-place scaling allows for precise adjustments to resource allocations in real time. This level of control enables the system to respond to fluctuations in demand with minimal overhead, ensuring that each function has access to the resources it requires without over-provisioning.
\end{enumerate}

\begin{figure*}[ht!]
  \centering

  \begin{subfigure}[b]{0.48\textwidth}
    \includegraphics[width=\textwidth]{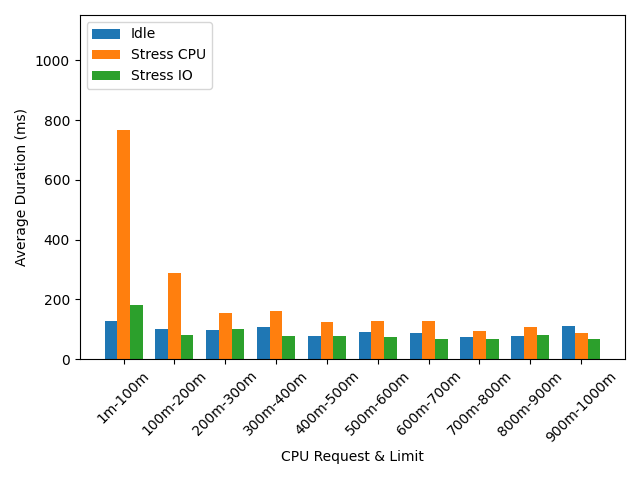}
    \caption{Incremental, Up, 1m, 1000m (Pattern, Direction, Initial, Target)}
    \label{fig:exp-size-100-a}
  \end{subfigure}
  ~
  \begin{subfigure}[b]{0.48\textwidth}
    \includegraphics[width=\textwidth]{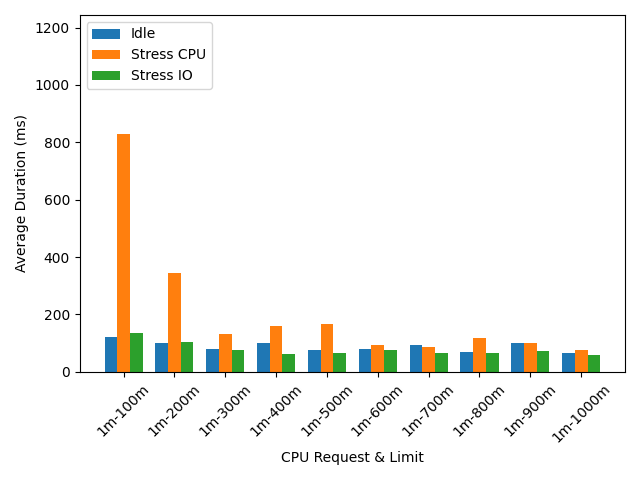}
    \caption{Cumulative, Up, 1m, 1000m}
    \label{fig:exp-size-100-b}
  \end{subfigure}

  \begin{subfigure}[b]{0.48\textwidth}
    \includegraphics[width=\textwidth]{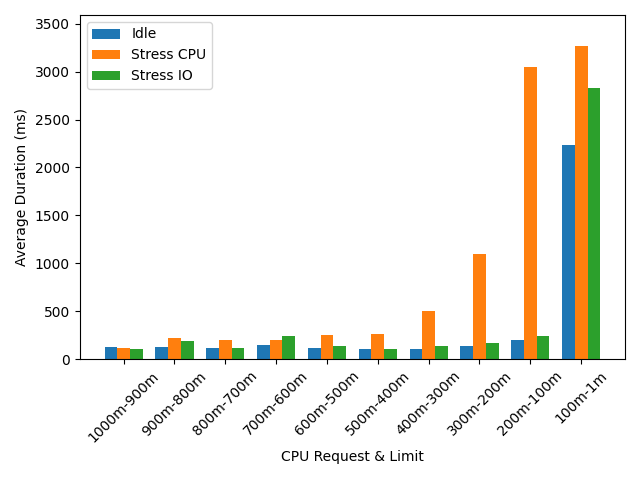}
    \caption{Incremental, Down, 1000m, 1m}
    \label{fig:exp-size-100-c}
  \end{subfigure}
  ~
  \begin{subfigure}[b]{0.48\textwidth}
    \includegraphics[width=\textwidth]{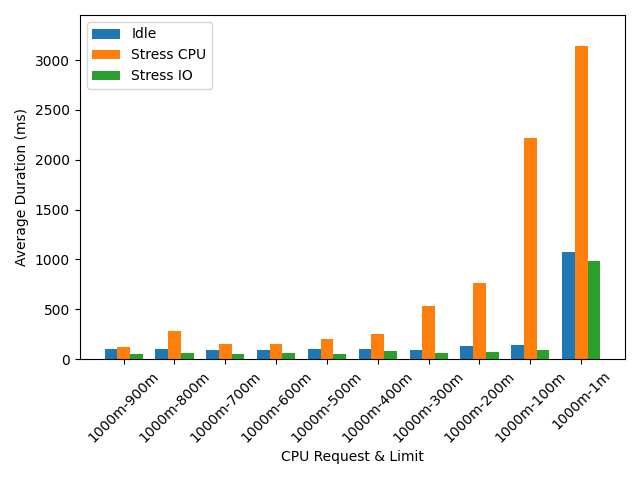}
    \caption{Cumulative, Down, 100m, 1m}
    \label{fig:exp-size-100-d}
  \end{subfigure}

  \caption{Avg Latency for Step Size = 100m}
  \label{fig:exp-size-100}
\end{figure*}

\section{Experiment}
\label{sec:exp}
In this section, we describe two experiments conducted for this paper. The first experiment is designed to assess the scaling overhead of the in-place scaling feature. The second experiment integrates this feature into a serverless platform. Both experiments were implemented using Go version 1.20. For our testing environment, we set up a local Kubernetes cluster using \textit{kind} (Kubernetes IN Docker), version 0.20.0, and used the kindest/node:v1.27.3 image to establish our node, which was allocated an 8-core CPU and 10GB of memory. We utilized Knative version 1.12 \cite{knative} for serverless integration and enabled the \textit{InPlacePodVerticalScaling} feature gate to allow the use of the in-place scaling feature \cite{inplace}.

\subsection{In-place Scaling Overhead}
To monitor the scaling duration, we utilized a single container and executed (\textit{exec}) into it to directly observe its control groups (cgroups). The duration was measured from the time the patch request was dispatched to the point when specified changes were detected within the cpu.max file in the cgroup directory. We recorded the timestamps of these actions and calculated the duration accordingly.

\begin{table}[h!]
    \centering
    \caption{Experiments for measuring in-place scaling duration}
    \begin{tabular}{ccccc}
        \toprule
        \textbf{Step Size} & \textbf{Pattern} & \textbf{Direction} & \textbf{Initial} & \textbf{Target} \\
        \midrule
        \multirow{4}{*}{100m} & \multirow{2}{*}{Incremental} & Up & 1m & 1000m \\
        \cline{3-5}
        & & Down & 1000m & 1m \\
        \cline{2-5}
        & \multirow{2}{*}{Cumulative} & Up & 1m & 1000m \\
        \cline{3-5}
        & & Down & 1000m & 1m \\
        \midrule
        \multirow{4}{*}{1000m} & \multirow{2}{*}{Incremental} & Up & 1m & 6000m \\
        \cline{3-5}
        & & Down & 6000m & 1m \\
        \cline{2-5}
        & \multirow{2}{*}{Cumulative} & Up & 1m & 6000m \\
        \cline{3-5}
        & & Down & 6000m & 1m \\
        \bottomrule
    \end{tabular}
    \label{table:exp}
\end{table}

\begin{figure*}[h!]
  \centering

  \begin{subfigure}[b]{0.48\textwidth}
    \includegraphics[width=\textwidth]{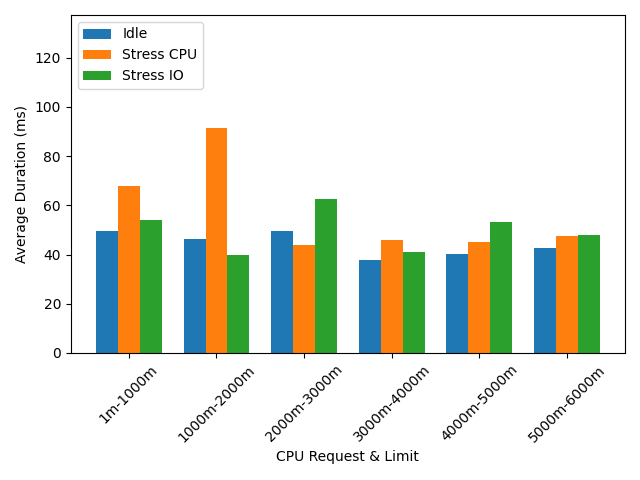}
    \caption{Incremental, Up, 1m, 6000m (Pattern, Direction, Initial, Target)}
    \label{fig:exp-size-1000-a}
  \end{subfigure}
  ~ 
  \begin{subfigure}[b]{0.48\textwidth}
    \includegraphics[width=\textwidth]{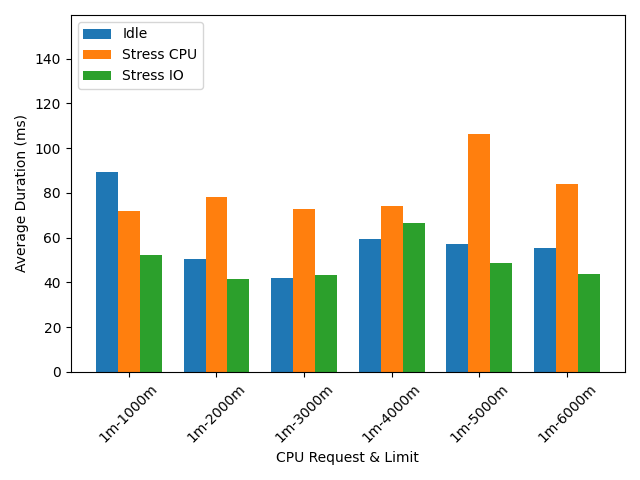}
    \caption{Cumulative, Up, 1m, 6000m}
    \label{fig:exp-size-1000-b}
  \end{subfigure}

  \begin{subfigure}[b]{0.48\textwidth}
    \includegraphics[width=\textwidth]{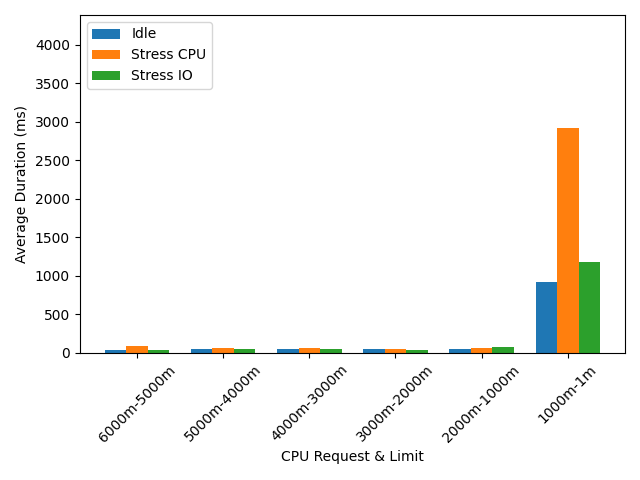}
    \caption{Incremental, Down, 6000m, 1m}
    \label{fig:exp-size-1000-c}
  \end{subfigure}
  ~ 
  \begin{subfigure}[b]{0.48\textwidth}
    \includegraphics[width=\textwidth]{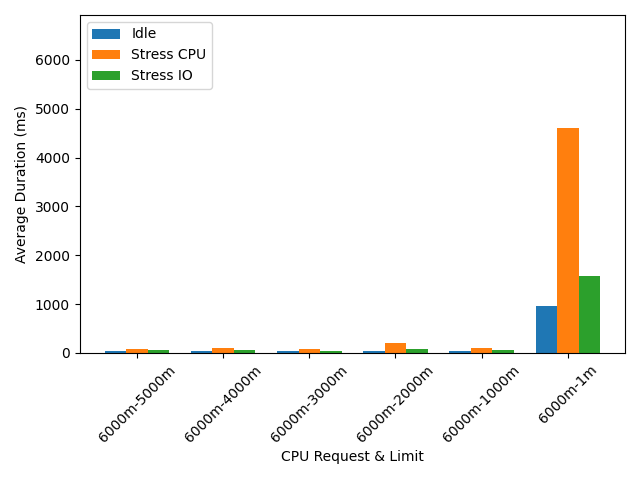}
    \caption{Cumulative, Down, 6000m, 1m}
    \label{fig:exp-size-1000-d}
  \end{subfigure}

  \caption{Avg Latency for Step Size = 1000m}
  \label{fig:exp-size-1000}
\end{figure*}

Table \ref{table:exp} outlines eight experimental configurations designed to assess the duration of in-place scaling. These experiments are divided into two scaling strategies: \textbf{Incremental} and \textbf{Cumulative}. The Incremental method indicates a stepwise increase, where each scaling operation builds upon the previous value. For example, in the first row of the Incremental pattern, the process begins by scaling from 1m (milliCPU) to 100m, then from 100m to 200m, and continues in this manner until it scales up to 1000m. In contrast, the Cumulative method involves scaling that continuously resets to the base value after each operation. For instance, in the first row of the Cumulative pattern, the process scales from 1m to 100m. It then resets back to 1m before scaling to 200m, repeating this pattern, cumulatively adding to the scale in each iteration until it reaches 1000m. Both methodologies were evaluated in both upward and downward directions, using two different intervals: one with finer granularity at 100m (0.1 CPU) and another with broader steps of 1000m (1 CPU).

Regarding workload type, we evaluated between the Idle and Busy states. The Idle state corresponds to a container that is operational but not executing any tasks, while the Busy state refers to a container actively processing tasks. To simulate the load under the Busy state, we used \texttt{stress-ng} \cite{stress-ng} as a stressor to generate both CPU and I/O stress on the system. Our evaluation results are categorized as follows:

\vspace{5pt}
\noindent
\textbf{Step Size = 100m, Up: } In Figure \ref{fig:exp-size-100-a} (Incremental) and \ref{fig:exp-size-100-b} (Cumulative), we observed that the scaling durations for the CPU workload are significantly longer than those for the Idle and Stress I/O, particularly within the first two intervals (1m-100m and 100m-200m). In the 1m-100m interval, the mean duration of the Stress CPU workload is 6.06 times longer in Figure \ref{fig:exp-size-100-a} and 6.83 times longer in Figure \ref{fig:exp-size-100-b} compared to the Idle state. For the 100m-200m interval, it is 2.88 times longer in Figure \ref{fig:exp-size-100-a} and 3.44 times longer in Figure \ref{fig:exp-size-100-b}. Although the differences in other intervals are not notable, the scaling durations for the Stress CPU workload are generally longer. 

\vspace{5pt}
\noindent
\textbf{Step Size = 100m, Down: } Figures \ref{fig:exp-size-100-c} (Incremental) and \ref{fig:exp-size-100-d} (Cumulative) also illustrates that performance is affected under CPU stress, as shown by the increasing trend when the target value decreases. While scaling up remains under 1 second, scaling down the CPU took up to 3.95 seconds for CPU workload. Fortunately, the impact of downward scaling on the CPU is deemed minimal for our purposes, as it occurs after the workload execution has been completed.

\vspace{5pt}
\noindent
\textbf{Step Size = 1000m, Up \& Down: } When employing a step size of 1000m (equivalent to 1 CPU), Figures \ref{fig:exp-size-1000-a} and \ref{fig:exp-size-1000-b} show minimal variations in latency across different workloads during both upward and downward scaling phases. The exception is noted in the last interval when scaling down from 1000m to 1m. These findings align with our previous experiments, suggesting that scaling down to 1m can lead to increased scaling overhead.

\begin{figure}[ht!]
  \centering

  \begin{subfigure}[b]{\columnwidth}
    \centering
    \includegraphics[width=\linewidth]{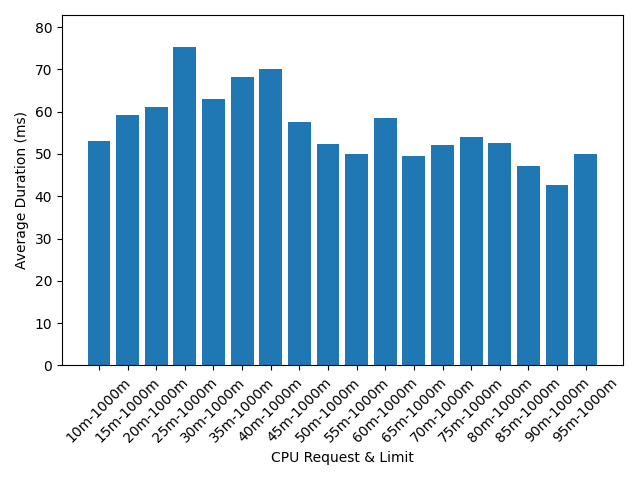}
    \caption{Increment from 5 milliCPU to 1000 milliCPU}
    \label{exp-5n-a}
  \end{subfigure}

  \begin{subfigure}[b]{\columnwidth}
    \centering
    \includegraphics[width=\linewidth]{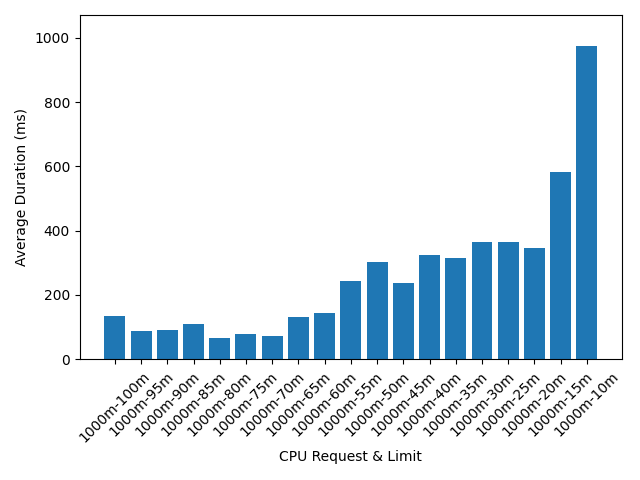} 
    \caption{Decrement from 1000 milliCPU to 5 milliCPU}
    \label{exp-5n-b}
  \end{subfigure}

  \caption{Average Latency of in-place scaling with finer intervals}
  \label{fig:exp-5n}
\end{figure}

The outcomes presented in Figure \ref{fig:exp-size-100} and Figure \ref{fig:exp-size-1000} highlight a disparity in the in-place scaling durations between the extremes of the scale, specifically between 1 milliCPU to 100 milliCPU and 1000 milliCPU to 1 milliCPU. This observation led us to conduct a more granular analysis. We dissected the 100 milliCPU bracket into smaller intervals of 5 milliCPU and evaluated it under idle conditions. As shown in Figure \ref{exp-5n-a}, this granular approach yielded a negligible variance in average latency for scaling up to 1000 milliCPU, with a mean duration of 56.44 ms and a standard deviation of 8.53 ms, indicating a consistent scaling duration regardless of the initial milliCPU values. On the other hand, Figure \ref{exp-5n-b} reveals an increasing scaling duration as the target milliCPU value diminishes. Nevertheless, in line with our previous observations, we consider this increase in duration during downscaling to be less important.

\subsection{Integrate In-place scaling into Serverless}
This section compares the cold start latency under three different scheduling policies: \textit{Cold}, \textit{Warm}, and \textit{In-place}. We utilized Knative \cite{knative}, an platform-agnostic serverless platform that offers scale-to-zero, and other features on Kubernetes, to conduct our experiments.

\begin{table}[h!]
  \centering
   \caption{Runtime measurements of various tests with 1 CPU}
    \begin{tabular}{ccc}
    \hline
    \textbf{Workload} & \textbf{Definition} & \multicolumn{1}{l}{\textbf{Runtime (ms)}} \\ \hline
    helloworld & return the “helloworld” string & 5.31 \\
    cpu & complicate math problem & 2465.18 \\
    io & open file n times & 2258.22 \\
    videos (10s) & ffmpeg watermark & 1659.03 \\
    videos (1m) & ffmpeg watermark & 13888.03 \\
    videos (10m) & ffmpeg watermark & 119028.34 \\ \hline
    \end{tabular}
  \label{tab:runtime_measurements}
\end{table}

For the cold policy, representative of the standard approach in serverless computing, we set the \textit{stable-window} parameter in Knative to the minimum duration of 6 seconds, which is the threshold for scaling down to zero in the absence of traffic (the default is 30 seconds). This configuration would not influence our test outcomes but it can allow for more efficient execution of the tests. In the warm scenario, we wanted to keep function instances always ready for handling the request, so we adjusted the \textit{min-scale} parameter to 1 to maintain at least one container in readiness at all times, thus eliminating cold start step. Regarding in-place scaling, we modified the queue-proxy in Knative, which is a sidecar container responsible for routing requests to the specific function container. These changes include adding a layer before the queue-proxy redirects the request, to allocate (1000m CPU in this study), and another layer after the request has been processed to deallocate (1m CPU in this study). To generate load, we used \texttt{k6} \cite{k6} to execute workloads detailed in Table \ref{tab:runtime_measurements}. All workloads were written in Python, including adding watermarks to videos from SeBS \cite{sebs}, and we measure the request times for each workload with 1 CPU.

\begin{figure}[h!]
  \centering
    \includegraphics[width=\linewidth]{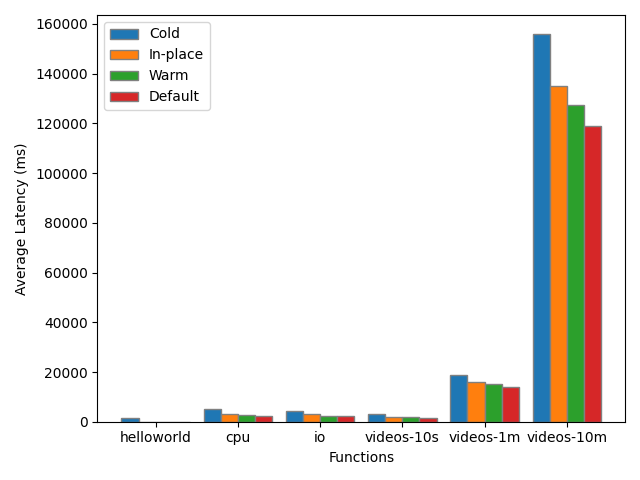}
  \caption{Average Latency of different scheduling policies}
  \label{fig:summary}
\end{figure}

Figure \ref{fig:summary} shows the outcomes using the \textit{Default} as a baseline to normalize each group's latencies. Detailed normalized results are provided in Table \ref{tab:normalize_summary}. We observed that the \textit{Warm} most closely mirrors the \textit{Default} in performance, as it is kept running continuously and is ready to serve, followed by \textit{In-place}. Comparing \textit{In-place} with \textit{Cold}, we see a decrease in latency. The in-place policy, while not as fast as the \textit{Warm} or \textit{Default}, still offers a considerable improvement over Cold starts. In functions like ``helloworld'', the improvement is by a factor of approximately 18.15 times, which highlights the effectiveness of in-place scaling in reducing latency by maintaining an environment that is quicker to activate than a full cold start. This approach provides a strategic advantage in managing resources more efficiently while still offering reduced latency, making it an alternative to the always-on warm policy.

\begin{table}[h]
\centering
\caption{Detailed Relative Latency of Figure \ref{fig:relative}}
\label{tab:normalize_summary}
\begin{tabular}{ccccc}
\toprule
  Function & Cold & In-place & Warm & Default \\
\midrule
helloworld &    286.99 &         15.81 &     3.87 &     1.00 \\
       cpu &      2.00 &          1.31 &     1.13 &     1.00 \\
        io &      1.89 &          1.46 &     1.09 &     1.00 \\
videos-10s &      1.88 &          1.24 &     1.03 &     1.00 \\
 videos-1m &      1.34 &          1.16 &     1.08 &     1.00 \\
videos-10m &      1.31 &          1.13 &     1.07 &     1.00 \\
\bottomrule
\end{tabular}
\end{table}

\begin{figure}[h!]
  \centering
    \includegraphics[width=\linewidth]{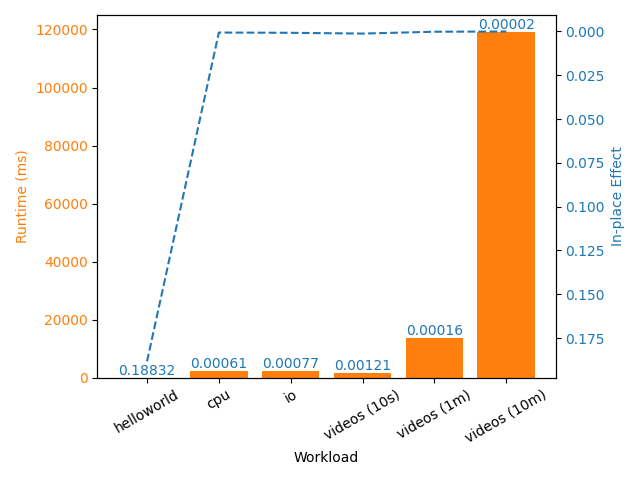}
  \caption{Runtime vs In-place Effect}
  \label{fig:relative}
\end{figure}

A detailed analysis of the \textit{In-place} policy reveals that relative durations tend to be shorter when the default runtime in Table \ref{tab:runtime_measurements} is longer. This inverse relationship is further depicted in Figure \ref{fig:relative}.

\section{Related Work}
\label{sec:related}
RUBAS \cite{rubas} employs Checkpoint/Restore in Userspace (CRIU) to resume container execution and migrate workloads, while COCOS \cite{cocos} utilizes Docker out of Docker (DooD) for similar purposes. However, both rely on external components for vertical scaling, whereas our method leverages the native Kubernetes in-place scaling feature. KOSMOS \cite{kosmos} introduces an autoscaling system that combines both vertical and horizontal scaling, but it does not assess its performance in a serverless context.

Only a few solutions have considered vertical scaling within the serverless paradigm specifically. Zhao et al. have developed tiny autoscalers \cite{tiny-autoscalers} that dynamically allocate CPU resources for serverless functions. However, since in-place updates were only a proposed concept then, they disabled the update mechanism. Therefore, to the best of our knowledge, our paper is the first to integrate the in-place scaling feature with serverless computing. 

\section{Conclusions and Future Work} 
\label{sec:conclusion}
In this work, we integrated the in-place scaling feature into serverless computing. Our experiments across various workloads revealed that the overhead associated with it is minimal. More importantly, our results demonstrate potential benefits in mitigating cold start issues, which are a significant concern in serverless architectures.

Looking ahead, our future work will dive into the complexities surrounding memory management and its consequential effects on cluster placement strategies. With the in-place scaling feature, we aim to evolve a holistic model that encapsulates both vertical and horizontal scaling dimensions. This exploration is expected to yield deeper insights and potential solutions that can enhance the performance and resource utilization in serverless computing.

\bibliographystyle{ACM-Reference-Format}
\bibliography{main}

\end{document}